# Functional Mediation Analysis with an Application to Functional Magnetic Resonance Imaging Data

Yi Zhao

Department of Biostatistics, Johns Hopkins School of Public Health

and

Xi Luo

Department of Biostatistics, School of Public Health, Brown University

and

Martin Lindquist

Department of Biostatistics, Johns Hopkins School of Public Health

and

Brian Caffo

Department of Biostatistics, Johns Hopkins School of Public Health

May 16, 2018

### Abstract

Causal mediation analysis is widely utilized to separate the causal effect of treatment into its direct effect on the outcome and its indirect effect through an intermediate variable (the mediator). In this study we introduce a functional mediation analysis framework in which the three key variables, the treatment, mediator, and outcome, are all continuous functions. With functional measures, causal assumptions and interpretations are not immediately well-defined. Motivated by a functional magnetic resonance imaging (fMRI) study, we propose two functional mediation models based on the influence of the mediator: (1) a concurrent mediation model and (2) a historical mediation model. We further discuss causal assumptions, and elucidate causal interpretations. Our proposed models enable the estimation of individual causal effect curves, where both the direct and indirect effects vary across time. Applied to a task-based fMRI study, we illustrate how our functional mediation framework provides a new perspective for studying dynamic brain connectivity. The R package cfma is available on CRAN.

*Keywords:* Structural equation model; Functional data analysis; Time-varying causal trajectory; Dynamic brain connectivity

# 1 Introduction

Causal mediation analysis is commonly used to separate the causal effect of a treatment into its direct effect on the outcome and its indirect effect through an intermediate variable (the mediator). Methods for performing causal mediation analysis on univariate measurement data have been extensively studied in recent years (Baron and Kenny, 1986; MacKinnon, 2008; Holland, 1988; Robins and Greenland, 1992; Pearl, 2001; Imai et al., 2010; VanderWeele, 2015). For time-dependent mediators and outcomes, existing studies have primarily focused on sparse longitudinal data (Avin et al., 2005; van der Laan and Petersen, 2008; VanderWeele, 2009; Goldsmith et al., 2016; Bind et al., 2016; Chen et al., 2016; Zheng and van der Laan, 2017; VanderWeele and Tchetgen Tchetgen, 2017). Recently, causal mediation analysis in high-dimensional settings have been explored (Huang and Pan, 2016; Chén et al., 2017).

In the neuroimaging context, causal mediation analysis is becoming an increasingly important method for assessing the intermediate effects of brain function on cognitive behavior (Wager et al., 2008, 2009b,a; Atlas et al., 2010, 2014; Woo et al., 2015). Current methodology focuses on either a single mediator or low-dimensional mediators with scalar measures. High-dimensional imaging based mediators were considered in Caffo et al. (2007), though the approach employed feature extraction with univariate mediation measures. For data measured in finer grids, Lindquist (2012) introduced the concept of functional mediation analysis, where the intermediate variable is a continuous function consisting of blood-oxygen-level dependent (BOLD) signal collected in a task-based functional magnetic resonance imaging (fMRI). Here the treatment was temperature and the outcome is self-reported pain scores, both scalar measures. Finally, in a recent study, Zhao and Luo (2017) introduced a framework integrating causal mediation with Granger causality for fMRI time series to capture the spatio-temporal dependencies and articulate brain causal mechanisms.

In this study, we extend the functional mediation concept to the scenario where the treatment, the mediator and the outcome are all continuous functions of time. A conceptual causal diagram is presented in Figure 1. This type of data arises frequently in medical, public health and biological research where multiple measurements are taken over time. Our approach extends methods from the area of functional data analysis (FDA), which is a collection of techniques (e.g., ANOVA and regression) to analyze data that take the form of functions (Ramsay, 2006), to the mediation setting. Our work builds on functional regression where both the response and covariates are functions. In this setting, there are currently three major types of models in use: (1) concurrent, (2) short-term, and (3) historical (Ramsay, 2006; Wang et al., 2016). The short-term and historical models can be represented using the same formulation. Thus, we consider them as a single type of functional regression model. Therefore, two types of functional mediation models will be introduced and causal estimands and identification assumptions associated with each will be studied.

The proposed approach will be applied to fMRI data, which is a major non-invasive tool for inferring brain connectivity. Recently, study that has focused on capturing time-varying brain connectivity is growing rapidly. Calhoun et al. (2014) introduced the concept of "chronnectome" to "describe metrics that allow a dynamic view of coupling". Current chronnectome research focuses on dynamic functional connectivity (the undirected association between brain regions) under both resting state (Chang and Glover, 2010; Cribben et al., 2012; Calhoun et al., 2013; Leonardi et al., 2013; Kucyi and Davis, 2014; Lindquist et al., 2014; Zalesky et al., 2014; Allen et al., 2014; Damaraju et al., 2014) and cognitive tasks (Sakoğlu et al. (2010); Warnick et al. (2017); Gonzalez-Castillo and Bandettini (2017)). Studies on time-varying effective brain connectivity (the directed association be-tween brain regions) are relatively scarce. To infer effective connectivity, dynamic causal modeling (Friston et al., 2003), dynamic directional models (Zhang et al., 2015, 2017), structural equation modeling and Granger causality in the form of vector autoregressive models are the commonly applied approaches (Lindquist, 2008). Samdin et al. (2015) proposed a vector autoregressive approach to estimate dynamic effective connectivity in an alternating resting-task block design. In this study, motivated by a response conflict task fMRI experiment, in which the participants perform motor responses to randomized STOP/GO stimuli, we investigate the dynamic intermediate effect of brain activities in the presupplementary motor area (preSMA) on activities in the primary motor cortex (M1) using proposed functional mediation framework.

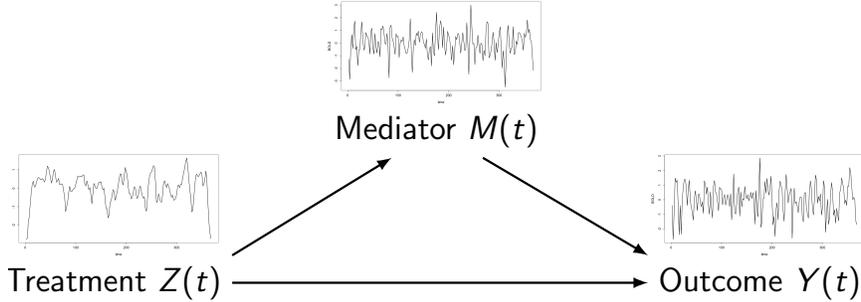

Figure 1: Conceptual causal diagram with functional treatment, mediator and outcome.

This paper is organized as follows. Section 2 introduces the two types of functional mediation models, and formulates causal assumptions and causal interpretations. In Section 3, we briefly present the methods of estimating model coefficient curves as well as causal estimands. Section 4 demonstrates the performance of the two types of models through simulations. We apply the proposed models on a task-based fMRI study and characterize the dynamic causal mechanisms in Section 5. Section 6 summarizes this paper with discussions and future directions.

## 2 Functional Mediation Models

In this section, we introduce two types of functional mediation models, which we denote the concurrent and historical influence mediation model. Without loss of generality, we assume that the data curves are centered and drop the intercept terms. Both models are generalizations of linear structural equation models (SEMs).

(1) The concurrent mediation model

   The concurrent model assumes SEM relationships hold for each time point and this is effectively a point-wise SEM. The model can be expressed as follows:

$$M(t) = Z(t)\alpha(t) + \epsilon_1(t), \qquad (1)$$

$$Y(t) = Z(t)\gamma(t) + M(t)\beta(t) + \epsilon_2(t), \qquad (2)$$

   where $\alpha(t)$, $\beta(t)$ and $\gamma(t)$ are coefficient curves ($t \in [0, T]$, $T \in \mathbb{R}^+$); and $\epsilon_1(t)$ and $\epsilon_2(t)$ are model error curves with mean zero.

(2) The historical influence mediation model

   The historical influence model models the accumulative effect of the dependent variables on the independent variable over the history $\Omega_t$. Extended to mediation analysis, the two regression models can be written as:

$$M(t) = \int_{\Omega_t^1} Z(s)\alpha(s,t) \, \mathrm{d}s + \epsilon_1(t), \qquad (3)$$

$$Y(t) = \int_{\Omega_t^2} Z(s)\gamma(s,t) \, \mathrm{d}s + \int_{\Omega_t^3} M(s)\beta(s,t) \, \mathrm{d}s + \epsilon_2(t), \qquad (4)$$

   where $\alpha(s,t)$, $\gamma(s,t)$ and $\beta(s,t)$ indicate the impact of the corresponding independent variable at time $s \in [0, T]$ on the dependent variable at time $t \in [0, T]$. One class of the historical model considers a fixed short period of history, i.e., $\Omega_t = [(t-\delta) \vee 0, t]$, where $a \vee b = \max(a, b)$ and $\delta$ is a small positive number representing the width of time window considered. This type of model assumes that the covariates effect only endure for a short period of time. Another class considers the whole history of influence, i.e., when $\delta \in [T, +\infty]$. In this case, $\Omega_t = [0, t]$, which varies over time. To make it more flexible, the influence window width of the three components can differ and $\Omega_t^j = [(t-\delta_j) \vee 0, t]$ for $j = 1, 2, 3$.

The concurrent model can be considered as a special case of the historical model with $\delta = 0$, and bivariate functions $\alpha(s,t)$, $\beta(s,t)$ and $\gamma(s,t)$ degenerate into one-dimensional functions with one dimension as constant. However, as discussed in Section 3, the estimation method under the concurrent model can be largely simplified compared to the historical model. Thus, we here consider them as two separate types of models. In a task-based fMRI study, one motivation of using the historical influence model with a small constant $\delta$ is to study the accumulative causal effects within a short period, for example a 20-second time window which is the approximate time for the heamodynamic response function (HRF) to recover from a stimulus (Friston et al., 1994, 1998). The whole-history model allows us determine the aggregated impact of the signal since the beginning of data recording.

## 2.1 Causal assumptions and interpretations

Using the potential outcome framework (Rubin, 1978, 2005), we first formulate the causal estimands of interest, i.e., the indirect effect (IE) and the direct effect (DE) (also referred as the controlled direct effect (VanderWeele, 2011)). Let $Y(t; \{z(s), m(s)\}_{\mathcal{H}_t})$ denote the potential outcome of $Y$ at time $t$ when the history of treatment $Z$ and mediator $M$ are at the level $\{z(s)\}_{\mathcal{H}_t}$ and $\{m(s)\}_{\mathcal{H}_t}$, respectively, where $\mathcal{H}_t = [0, t]$; and $M(t; \{z(s)\}_{\mathcal{H}_t})$ denote the outcome of the mediator at time $t$ if $Z$ has the history $\{z(s)\}_{\mathcal{H}_t}$.

(1) Under the concurrent mediation model (1) and (2),

$$
\begin{aligned}
\text{IE}(t) &= \mathbb{E}\left[Y(t;\{z(s), m(s;\{z(u)\}_{\mathcal{H}_s})\}_{\mathcal{H}_t}) - Y(t;\{z(s), m(s;\{z'(u)\}_{\mathcal{H}_s})\}_{\mathcal{H}_t})\right] \\
&= [z(t) - z'(t)]\,\alpha(t)\beta(t),
\end{aligned} \tag{5}
$$

$$
\begin{aligned}
\text{DE}(t;\{m(s)\}_{\mathcal{H}_t}) &= \mathbb{E}\left[Y(t;\{z(s), m(s)\}_{\mathcal{H}_t}) - Y(t;\{z'(s), m(s)\}_{\mathcal{H}_t})\right] \\
&= [z(t) - z'(t)]\,\gamma(t).
\end{aligned} \tag{6}
$$

(2) Under the historical influence mediation model (3) and (4),

$$
\begin{aligned}
\text{IE}(t) &= \mathbb{E}\left[Y(t;\{z(s), m(s;\{z(u)\}_{\mathcal{H}_s})\}_{\mathcal{H}_t}) - Y(t;\{z(s), m(s;\{z'(u)\}_{\mathcal{H}_s})\}_{\mathcal{H}_t})\right] \\
&= \int_{\Omega_t^3} \left(\int_{\Omega_s^1} [z(u) - z'(u)]\,\alpha(u,s)\,\mathrm{d}u\right) \beta(s,t)\,\mathrm{d}s,
\end{aligned} \tag{7}
$$

$$
\begin{aligned}
\text{DE}(t;\{m(s)\}_{\mathcal{H}_t}) &= \mathbb{E}\left[Y(t;\{z(s), m(s)\}_{\mathcal{H}_t}) - Y(t;\{z'(s), m(s)\}_{\mathcal{H}_t})\right] \\
&= \int_{\Omega_t^2} [z(s) - z'(s)]\,\gamma(s,t)\,\mathrm{d}s.
\end{aligned} \tag{8}
$$

Under both models, the DE does not depend on the controlled level of the mediator. The concurrent model assumes the linear relationship holds at each time point, and therefore, discretizes the continuous functions. Thus, the causal estimands have the same formulation as in classic causal mediation results (VanderWeele, 2015). For the historical influence model, the interpretation of DE is straightforward and it reveals the integrated direct treatment effects over the time period $\Omega_t^2$. For example, the shaded area in Figure 2a under the historical influence mediation model. The calculation of IE is more complex. It first accounts for the cumulative treatment effect on the mediator and then further integrates the $Z \to M$ and $M \to Y$ path effects over the considered time period. Figure 2b demonstrates the double integral in (7) under the historical influence mediation model assuming $\delta_1 = \delta_2 = \delta_3 = \delta$. For given $t$, at time $s \in [t - \delta, t]$, $\tilde{\alpha}(s) = \int_{s-\delta}^{s}(z(u) - z'(u))\alpha(u, s) \, \mathrm{d}u$ first assembles the treatment effect on the mediator over time period $\Omega_s^1 = [s - \delta, s]$, and thus $\tilde{\alpha}(s)\beta(s, t)$ denotes the indirect effect at time $s$ on the outcome at time $t$. Integrating over $\Omega_t^3 = [t - \delta, t]$ yields the indirect effect at time $t$, i.e., the volume of the shaded area in Figure 2b. Here, we mainly discuss the causal estimands of two types of functional mediation models. In a more general scenario, model types of the mediator and the outcome may differ. We summarize the formulation of direct and indirect effects under some other types of functional mediation models in Table 1.

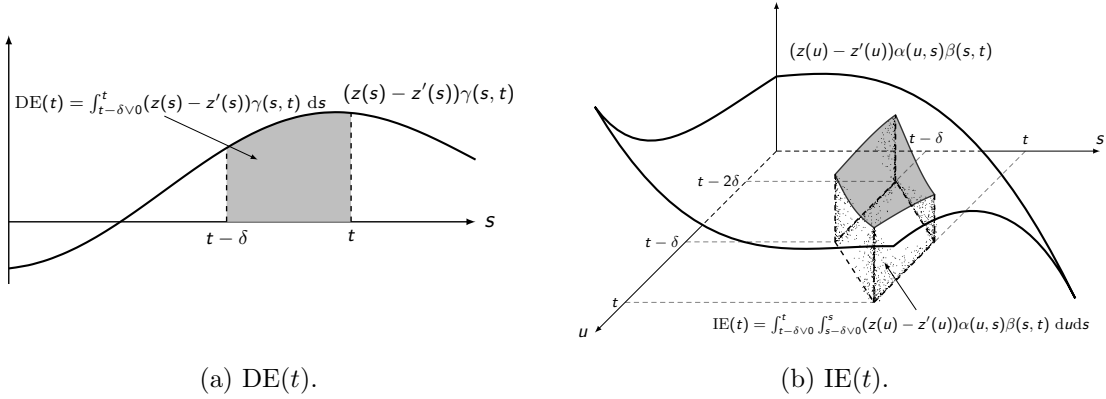

Figure 2: Direct and indirect effects at time $t$ under the historical influence mediation model (assuming $\delta_1 = \delta_2 = \delta_3 = \delta$).

To identify the direct/indirect effect, we impose the following causal assumptions. Let $\mathcal{O}_t = \left\{ \{Z(s)\}_{\mathcal{H}_t \setminus \{t\}}, \{M(s)\}_{\mathcal{H}_t \setminus \{t\}}, \{Y(s)\}_{\mathcal{H}_t \setminus \{t\}} \right\}$ the observed data up to time $t$.

Table 1: Causal estimands (the natural direct effect at time $t$ ($\mathrm{DE}(t)$) and the natural indirect effect at time $t$ ($\mathrm{IE}(t)$)) under various types of functional mediation models. C-CH represents the scenario with concurrent $M$-$Z$ model, concurrent $Y$-$Z$ and historical $Y$-$M$ in the outcome model. The rest are defined analogously.

|  |  | $\mathrm{DE}(t)$ |  |
| --- | --- | --- | --- |
|  |  | $[z(t) - z'(t)]\gamma(t)$ | $\int_{\Omega_t^2}[z(s) - z'(s)]\gamma(s,t)\,\mathrm{d}s$ |
| $\mathrm{IE}(t)$ | $[z(t) - z'(t)]\alpha(t)\beta(t)$ | C-CC | C-HC |
|  | $\int_{\Omega_t^3}[z(s) - z'(s)]\alpha(s)\beta(s,t)\,\mathrm{d}s$ | C-CH | C-HH |
|  | $\left(\int_{\Omega_t^1}[z(s) - z'(s)]\alpha(s)\,\mathrm{d}s\right)\beta(t)$ | H-CC | H-HC |
|  | $\int_{\Omega_t^3}\left(\int_{\Omega_s^1}[z(u) - z'(u)]\alpha(u,s)\,\mathrm{d}u\right)\beta(s,t)\,\mathrm{d}s$ | H-CH | H-HH |

**Assumption 1** There is no (unmeasured) "treatment-outcome confounder", i.e.,

$$Y(t; \{z(s), m(s)\}_{\mathcal{H}_t}) \perp\!\!\!\perp Z(t) \mid \mathcal{O}_t. \tag{9}$$

**Assumption 2** There is no (unmeasured) "treatment-mediator confounder", i.e.,

$$M(t; \{z(s)\}_{\mathcal{H}_t}) \perp\!\!\!\perp Z(t) \mid \mathcal{O}_t. \tag{10}$$

**Assumption 3** There is no (unmeasured) "mediator-outcome confounder", i.e.,

$$Y(t; \{z'(s), m(s)\}_{\mathcal{H}_t}) \perp\!\!\!\perp M(t; \{z(s)\}_{\mathcal{H}_t}) \mid Z(t), \mathcal{O}_t. \tag{11}$$

Assumptions 1-3 are extensions of the standard causal mediation assumptions (Vander-Weele, 2015) to the functional data scenario. Additionally, we assume the stable unit treatment value assumption (SUTVA, Rubin (1978, 1980)) is satisfied. From assumption 2, we have

$$\begin{aligned} \mathbb{E}\left[M(t;\{z(s)\}_{\mathcal{H}_t}) \mid \mathcal{O}_t\right] &= \mathbb{E}\left[M(t;\{z(s)\}_{\mathcal{H}_t}) \mid Z(t)=z(t), \mathcal{O}_t\right] \\ &= \mathbb{E}\left[M(t;\{Z(s)\}_{\mathcal{H}_t}) \mid Z(t)=z(t), \mathcal{O}_t\right]; \end{aligned}$$

and under assumptions 1-3,

$$\begin{aligned} &\mathbb{E}\left[Y(t;\{z(s), m(s,\{z(u)\}_{\mathcal{H}_t})\}_{\mathcal{H}_t}) \mid \mathcal{O}_t\right] \\ &= \mathbb{E}\left[Y(t;\{z(s), m(s,\{z(u)\}_{\mathcal{H}_t})\}_{\mathcal{H}_t}) \mid Z(t)=z(t), \mathcal{O}_t\right] \\ &= \mathbb{E}\left[Y(t;\{z(s), m(s,\{z(u)\}_{\mathcal{H}_t})\}_{\mathcal{H}_t}) \mid Z(t)=z(t), M(t;\{z(s)\}_{\mathcal{H}_t})=m(t;\{z(s)\}_{\mathcal{H}_t}), \mathcal{O}_t\right] \\ &= \mathbb{E}\left[Y(t;\{Z(s), M(s,\{Z(u)\}_{\mathcal{H}_t})\}_{\mathcal{H}_t}) \mid Z(t)=z(t), M(t;\{Z(s)\}_{\mathcal{H}_t})=m(t;\{z(s)\}_{\mathcal{H}_t}), \mathcal{O}_t\right]. \end{aligned}$$

Assuming that the mediation models are correctly specified, the causal estimands (DE and IE) can then be estimated from the observed data.

# 3 Methods

## 3.1 Estimation method

### 3.1.1 Concurrent mediation model

The concurrent mediation model (1) and (2) can be written in a more general form as

$$Y(t) = X(t)\theta(t) + \epsilon(t), \qquad (12)$$

where $Y(t) = (Y_1(t), \ldots, Y_N(t))^\top$ is the vector of observed dependent variable from $N$ subjects at time $t$, $t \in [0,T]$, $T \in \mathbb{R}^+$; $X(t)$ is the $N \times q$ design matrix; $\theta(t) = (\theta_1(t), \ldots, \theta_q(t))^\top$ is the coefficient curves of $q$ covariates; and $\epsilon(t) = (\epsilon_1(t), \ldots, \epsilon_N(t))^\top$ is a vector of $N$ zero-mean stochastic processes. Various approaches have been introduced to estimate the coefficient curves, for a review see Wang et al. (2016). In this study, we employ a one-step penalized least squares approach (Ramsay, 2006). Similar to ordinary least square regression, the aim is to minimize the $\ell_2$-loss

$$\text{SSE}(\theta) = \int_0^T \|Y(t) - X(t)\theta(t)\|_2^2 \, dt. \tag{13}$$

To control the smoothness of the estimates for $\theta_j$'s, a roughness penalty is considered,

$$\text{PEN}_j(\theta_j) = \lambda_j \int_0^T [\mathcal{L}_j \theta_j(t)]^2 \, dt, \tag{14}$$

where $\lambda_j$ is the tuning parameter which can be chosen through cross-validation (Ramsay, 2006; Wang et al., 2016; Lindquist, 2012) and $\mathcal{L}_j$ is a linear differential operator, such as the curvature operator $\mathcal{L}_j = \mathcal{D}^2$ or the harmonic acceleration operator $\mathcal{L}_j = \omega^2 \mathcal{D} + \mathcal{D}^3$ ($\mathcal{D}$ is the differential operator and $\omega$ is the angular frequency), $j = 1, \ldots, q$. The weighted regularized fitting criterion is given by

$$\text{LMSSE}(\theta) = \int_0^T \|Y(t) - X(t)\theta(t)\|_2^2 \, dt + \sum_{j=1}^q \lambda_j \int_0^T [\mathcal{L}_j \theta_j(t)]^2 \, dt. \tag{15}$$

Suppose each coefficient function $\theta_j(t)$ has an expansion of form

$$\theta_j(t) = \sum_{k=1}^{K_j} g_{kj} \phi_{kj}(t) = \boldsymbol{\phi}_j^\top(t) \mathbf{g}_j, \tag{16}$$

where $\phi_{kj}(t)$ is the basis function and $g_{kj}$ is the corresponding coefficient, $k = 1, \ldots, K_j$. Various basis systems can be used for function approximation. When the underlying function is periodic, a Fourier basis is well suited. Other basis systems, including polynomials, kernel functions and B-spline basis, are also commonly applied. Let $K_\theta = \sum_{j=1}^q K_j$ be the total number of basis functions. Define

$$\mathbf{g} = \begin{pmatrix} \mathbf{g}_1 \\ \vdots \\ \mathbf{g}_q \end{pmatrix}, \quad \mathbf{\Phi}(t) = \begin{pmatrix} \boldsymbol{\phi}_1^\top(t) & & \\ & \ddots & \\ & & \boldsymbol{\phi}_q^\top(t) \end{pmatrix}, \quad \mathbf{U} = \begin{pmatrix} U_1 & & \\ & \ddots & \\ & & U_q \end{pmatrix},$$

$$U_j = \lambda_j \int_0^T [\mathcal{L}_j \phi_j(t)][\mathcal{L}_j \phi_j^\top(t)] \, dt.$$

Model (12) can be expressed as

$$Y(t) = X(t)\mathbf{\Phi}(t)\mathbf{g} + \epsilon(t), \tag{17}$$

and the solution that minimizes criterion (15) is given by

$$\hat{\mathbf{g}} = \left[ \int_0^T \mathbf{\Phi}^\top(t) X^\top(t) X(t) \mathbf{\Phi}(t) \, dt + \mathbf{U} \right]^{-1} \left[ \int_0^T \mathbf{\Phi}^\top(t) X^\top(t) Y(t) \, dt \right]. \tag{18}$$

### 3.1.2 Historical mediation model

The general form of the historical mediation model (3) and (4) is

$$Y(t) = \int_{\Omega_t} X(s)\theta(s,t) \, ds + \epsilon(t). \tag{19}$$

In this type of model, the model coefficient $\theta(s,t)$ is a bivariate function. A double expansion in terms of $K_{1j}$ basis functions $\phi_{kj}$ with respect to $s$ and $K_{2j}$ basis functions $\eta_{lj}$ with respect to $t$ is employed, i.e.,

$$\theta_j(s,t) = \sum_{k=1}^{K_{1j}} \sum_{l=1}^{K_{2j}} g_{klj} \phi_{kj}(s)\eta_{lj}(t) = \boldsymbol{\phi}_j^\top(s)\mathbf{G}_j\boldsymbol{\eta}_j(t) = (\boldsymbol{\eta}_j^\top(t) \otimes \boldsymbol{\phi}_j^\top(s))\text{vec}(\mathbf{G}_j), \tag{20}$$

where $\mathbf{G}_j = (g_{klj})$ is a $K_{1j} \times K_{2j}$ matrix of coefficients, $j = 1, \ldots, q$; and $\otimes$ is the Kronecker product operator. Let

$$\mathbf{G} = \begin{pmatrix} \text{vec}(\mathbf{G}_1) \\ \vdots \\ \text{vec}(\mathbf{G}_q) \end{pmatrix}, \mathbf{D}(t) = \begin{pmatrix} \boldsymbol{\eta}_1^\top(t) \otimes X_1^*(t) & \cdots & \boldsymbol{\eta}_q^\top(t) \otimes X_q^*(t) \end{pmatrix}, X_j^*(t) = \int_{\Omega_t} X_j(s)\boldsymbol{\phi}_j^\top(s) \, ds.$$

Model (19) is then rewritten as

$$Y(t) = \mathbf{D}(t)\mathbf{G} + \epsilon(t). \tag{21}$$

Two roughness penalty functions should be utilized to control the smoothness of the bivariate function $\theta(s,t)$. With respect to $s$,

$$\begin{aligned}
\text{PEN}_s(\theta) &= \int_0^T \int_0^T [\mathcal{L}_s\theta(s,t)][\mathcal{L}_s\theta^\top(s,t)] \,\mathrm{d}s\mathrm{d}t \\
&= \mathbf{G}^\top \text{diag}\left\{\int_0^T \boldsymbol{\eta}_j(t)\boldsymbol{\eta}_j^\top(t) \,\mathrm{d}t \otimes \int_0^T \mathcal{L}_s\boldsymbol{\phi}_j(s)\mathcal{L}_s\boldsymbol{\phi}_j^\top(s) \,\mathrm{d}s\right\} \mathbf{G} \\
&\triangleq \mathbf{G}^\top \mathbf{U} \mathbf{G};
\end{aligned} \tag{22}$$

and with respect to $t$

$$\begin{aligned}
\text{PEN}_t(\theta) &= \int_0^T \int_0^T [\mathcal{L}_t\theta(s,t)][\mathcal{L}_t\theta^\top(s,t)] \,\mathrm{d}s\mathrm{d}t \\
&= \mathbf{G}^\top \text{diag}\left\{\int_0^T \mathcal{L}_t\boldsymbol{\eta}_j(t)\mathcal{L}_t\boldsymbol{\eta}_j^\top(t) \,\mathrm{d}t \otimes \int_0^T \boldsymbol{\phi}_j(s)\boldsymbol{\phi}_j^\top(s) \,\mathrm{d}s\right\} \mathbf{G} \\
&\triangleq \mathbf{G}^\top \mathbf{V} \mathbf{G}.
\end{aligned} \tag{23}$$

Minimizing the $\ell_2$-loss along with the two roughness penalties,

$$\text{LMSSE}(\theta) = \int_0^T \|Y(t) - \mathbf{D}(t)\mathbf{G}\|_2^2 \,\mathrm{d}t + \lambda_s \text{PEN}_s(\theta) + \lambda_t \text{PEN}_t(\theta), \tag{24}$$

where $\lambda_s$ and $\lambda_t$ are tuning parameters, the solution for $\mathbf{G}$ is

$$\hat{\mathbf{G}} = \left[\int_0^T \mathbf{D}^\top(t)\mathbf{D}(t) \,\mathrm{d}t + \lambda_s \mathbf{U} + \lambda_t \mathbf{V}\right]^{-1} \left[\int_0^T \mathbf{D}^\top(t)Y(t) \,\mathrm{d}t\right]. \tag{25}$$

## 3.2 Inference

As the asymptotic variance expression in functional regression is not straight forward to compute, we propose a subject-level bootstrapping procedure to obtain the point-wise confidence bands of the estimated curves. We assume the concurrent mediation model is the true underlying causal mechanism as an example. For the $b$th bootstrap sample $(Z^{(b)}(t), M^{(b)}(t), Y^{(b)}(t))$:

(i) Attain model parameter curves $\hat{\alpha}^{(b)}$, $\hat{\beta}^{(b)}$ and $\hat{\gamma}^{(b)}$ using the methods described in Sections 3.1.1.

(ii) For subject $i$, plug in the estimated coefficient curves into (5) and (6) to yield an estimate of the indirect effect $\text{IE}_i^{(b)}(t)$ and the direct effect $\text{DE}_i^{(b)}(t)$, respectively.

Repeat procedures (i) and (ii) $B$ times. Point-wise confidence bands can be then calculated using either the percentile or bias-corrected approach (Efron, 1987).

# 4 Simulation Study

In the simulation study, we consider generating data from both types of models and compare the performance with a classic mediation analysis approach (Baron and Kenny, 1986). As we aim to simulate a realistic task-based fMRI study, we propose two competitive approaches: (1) a multilevel mediation approach (Kenny et al., 2003) directly on the functional observations, where for each subject the Baron and Kenny approach is applied directly on the data which discretizes the continuous time and assumes each time point as a randomized trial; and (2) a multilevel mediation approach based on the single-trial activations, where at the subject level, single-beta activation are first extracted using a general linear model (Duann et al., 2002) and then the mediation analysis is conducted on the beta coefficients. This is a similar approach as that described in Atlas et al. (2010) but with both the mediator and outcome being brain activation.

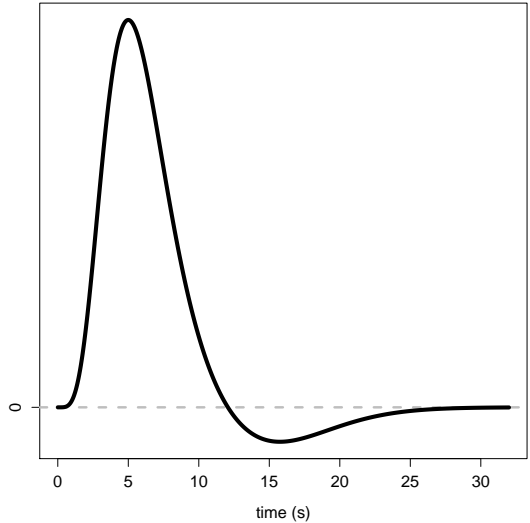

Figure 3: The canonical HRF used in the simulation study.

We simulate the treatment function $Z(t)$ as the convolution of a series of event times (with 40 seconds inter-trial interval) and the canonical hemodynamic response function (HRF) (Figure 3). The event condition is randomly assigned to be a "case" or "control" event, using a Bernoulli distribution with probability 0.5. Under the concurrent mediation model, $\alpha(t) = \sin(2\pi t/T)$, $\beta(t) = \cos(2\pi t/T) - t/T$, and $\gamma(t) = -\sin(2\pi t/T)$; and under the historical influence mediation model (with $\delta = 6$), $\alpha(s,t) = \sin(2\pi(s+t)/(2T)) + (s-t)/(2T)$, $\beta(s,t) = \cos(2\pi(s-t)/(2T)) - (s+t)/(2T)$, and $\gamma(s,t) = -\sin(2\pi(s+t)/(2T)) + (s-t)/(2T)$. The standard error of the error curves are set to be one. We generate $N = 50$ observations for each data generating mechanism. To simulate an fMRI study, we generate 150 data points with TR = 2 s (time range $[0, 300]$). Simulation studies are repeated 200 times.

Table 2 presents the results from the multilevel mediation method on either the time courses (KKB) or the the extracted single-beta activations (beta-KKB). Both methods attain a static estimate of the causal effects over the whole time period. With $Z(t) = 1$ and $Z'(t) = 0$, under the concurrent model, the average of IE$(t)$ is 0.158 and the average of DE$(t)$ 0; under the historical model, the average of IE$(t)$ is 5.602 and the average of DE$(t)$ -0.029. The beta-KKB approach yields a good estimate of the average direct effect under the concurrent model, which summarizes the time-varying direct effect of each trial into an average. The estimate of the average indirect effect is off since in general, the average of the product differs from the product of the average. The estimates from the KKB approach diverge from the truth.

Table 2: Estimate of direct effect (DE) and indirect effect (IE) using the multilevel mediation (KKB) and multilevel mediation on beta-activation (beta-KKB) approaches for data generated from both the concurrent model and the historical influence model ($\delta = 6$). Power is calculated from 500 bootstrap samples in each replication.

|  |  | Concurrent | | Historical | |
| --- | --- | --- | --- | --- | --- |
|  |  | Estimate (SE) | Power | Estimate (SE) | Power |
| KKB | IE | 0.017 (0.004) | 0.976 | 0.014 (0.001) | 1.000 |
|  | DE | 0.158 (0.004) | 1.000 | -0.937 (0.007) | 1.000 |
| beta-KKB | IE | -0.009 (0.072) | 0.049 | -0.022 (0.161) | 0.049 |
|  | DE | -0.000 (0.028) | 0.055 | 0.001 (0.039) | 0.067 |

For our proposed functional mediation methods, we choose Fourier basis of order five and the curvature operator (second derivative) in the roughness penalty to estimate model coefficient trajectories. The smoothing parameter $\lambda$'s are chosen using five-fold cross-validation. To evaluate the performance of the functional estimates, we define the mean squared error (MSE) for the subject-level causal curves as

$$\text{MSE}(\hat{\theta}) = \frac{1}{N} \sum_{i=1}^{N} \int_0^T \left[ \hat{\theta}_i(t) - \theta_i(t) \right]^2 \, dt, \tag{26}$$

where $\theta_i$ can be either the individual direct or indirect effect; and the mean absolute error (MAE) as

$$\text{MAE}(\hat{\theta}) = \frac{1}{N} \sum_{i=1}^{N} \int_0^T |\hat{\theta}_i(t) - \theta_i(t)| \, dt. \tag{27}$$

For population-level parameter curves, for example the causal curves when $Z_{i1}(t) = Z_1(t)$ and $Z_{i0}(t) = Z_0(t)$ where all the subjects in the same arm receive the same treatment trajectory, the MSE and MAE are defined the same as in (26) and (27) with $\theta_i(t) = \theta(t)$. In addition, we introduce the definition of bias as

$$\text{Bias}(\hat{\theta}) = \int_0^T |\hat{\theta}(t) - \theta(t)| \, dt, \tag{28}$$

where for the direct and indirect effects $\hat{\theta}(t) = \sum_{i=1}^{N} \hat{\theta}_i(t)/N$, and for model parameters, for example $\alpha(t)$ under the concurrent model, $\hat{\alpha}(t)$ is the estimate obtained using the method introduced in Section 3.1.1. We compare the performance of the proposed functional mediation model with the multilevel mediation approaches in Table 3. For both the concurrent model and the historical model, when the model is correctly specified, the functional mediation approach achieves good estimates of the causal curves. Figure 4 shows the estimates under the concurrent model and Figure 5 under the historical model. The estimated curves are very close to the true causal trajectories. Figure A.1 in Supplementary Section A compares the mean squared prediction error (MSPE) of $M$ and $Y$ under different $\delta$ choices when the true model is the historical mediation model with $\delta = 6$, where the MSPE is calculated using formula (26) by replacing $\theta$ with $M$ and $Y$, respectively. The lowest MSPE of both $M$ and $Y$ are achieved when $\delta = 6$. This suggests that one can choose the influence window width $\delta$ by comparing the MSPE, which is an analogous strategy of choosing the order of autoregressive model in time series analysis by minimizing the forecast mean squared error (Akaike, 1969; Lütkepohl, 2005).

Table 3: Mean squared error (MSE) and mean absolute error (MAE) of subject-specific causal curves, as well as the bias of causal curves when $Z(t) = 1$ and $Z'(t) = 0$.

| Model | Method | IE | | | DE | | |
|---|---|---|---|---|---|---|---|
| | | MSE | MAE | Bias | MSE | MAE | Bias |
| Concurrent | Concurrent | 11.712 | 24.575 | 11.188 | 9.597 | 23.498 | 10.955 |
| | KKB | 1258.081 | 258.827 | 129.203 | 2353.447 | 396.842 | 193.376 |
| | beta-KKB | 1317.486 | 267.940 | 131.813 | 2247.688 | 386.386 | 190.959 |
| Historical | Historical | 331.504 | 185.216 | 91.848 | 256.857 | 150.777 | 74.247 |
| | KKB | $3.0 \times 10^5$ | 5547.790 | 3348.242 | $3.8 \times 10^4$ | 2121.008 | 1158.878 |
| | beta-KKB | $3.0 \times 10^5$ | 5560.168 | 3348.566 | $3.5 \times 10^4$ | 2025.267 | 1145.337 |

## 5 A Functional MRI Study

We apply the proposed functional mediation models to a data set downloaded from the OpenfMRI database (accession number ds000030). The experiment consisted of $N = 121$ healthy right-handed participants. The participants were asked to perform a response conflict task, where the conflict occurs between the GO trial (press button seeing the GO stimulus) and the STOP trial (abstaining from pressing when a STOP signal, a 500 Hz tone, is presented through headphones after the GO stimulus). The STOP/GO stimuli were randomly intermixed with 96 GO trials and 32 STOP trials, at randomly jittered

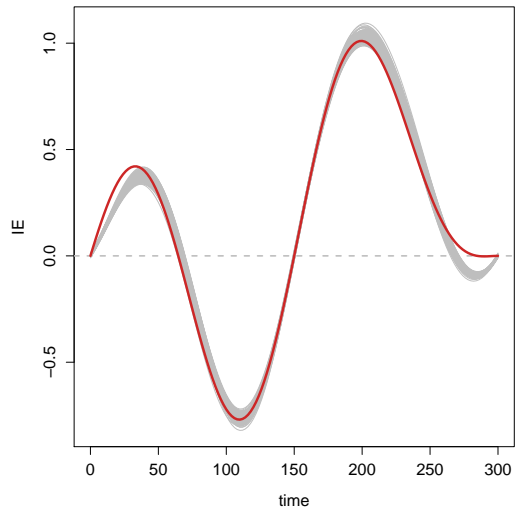
(a) IE

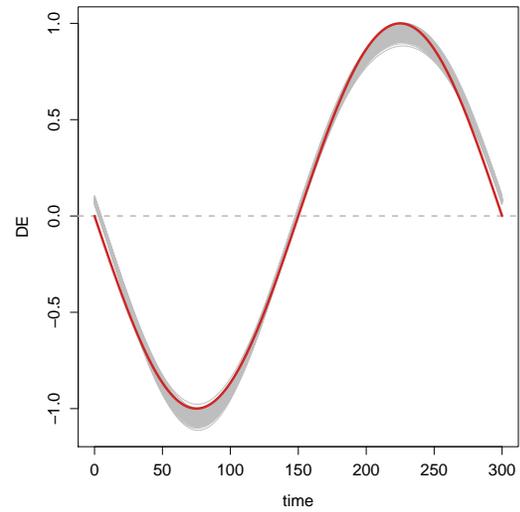
(b) DE

Figure 4: The estimate of natural direct and indirect effect curves (with $Z(t) = 1$ and $Z'(t) = 0$) under the concurrent model. The gray curves are the estimates of 200 replicates and the red curve is the truth.

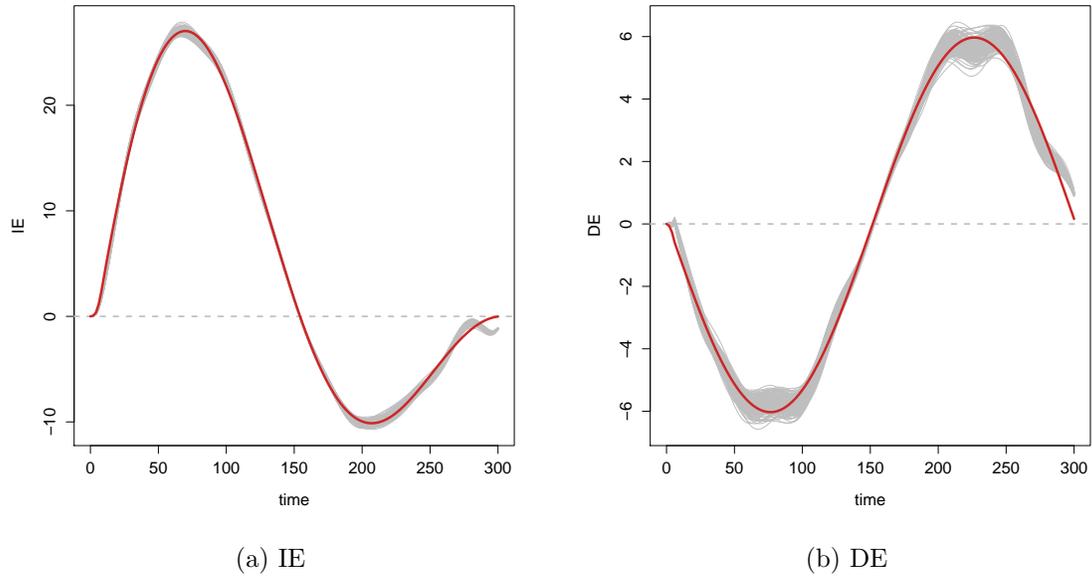

Figure 5: The estimate of natural direct and indirect effect curves (with $Z(t) = 1$ and $Z'(t) = 0$) under the historical model with $\delta = 6$. The gray curves are the estimates of 200 replicates and the red curve is the truth.

time intervals. For details about the experiment, see Poldrack et al. (2016). The objective is to quantify the time-varying intermediate role of preSMA on the functioning of M1. The causal direction, i.e., preSMA $\to$ M1, has previously been verified in Duann et al. (2009) and Obeso et al. (2013).

Preprocessing was performed using Statistical Parametric Mapping version 5 (SPM5) (Wellcome Department of Imaging Neuroscience, University of College London, UK) for both anatomical and functional images. It consists of slice timing correction, realignment, coregistration, normalization and smoothing steps. Our study interest focuses on the causal dynamics between the mediator brain region preSMA (MNI coordinate: (-4,-8,60)) and the outcome brain region M1 (MNI coordinate: (-41,-20,62)). BOLD time series consisting of $n_i = n = 184$ time points (TR = 2 s) were extracted from these two brain regions by averaging over voxels within a 10 mm radius sphere. For each participant, the treatment function was generated by convolving the stimulus onsets with the canonical HRF (Friston, 2009).

We compare the estimated causal curves from the proposed functional mediation models for STOP and GO trials separately. For each trial condition, the signal induced by the other condition is removed from the general linear model. We assume that *i)* there is no unmeasured confounding factor that have an impact on the BOLD signals of the two brain regions; and *ii)* the HRF applied to preSMA and M1 are the same. For the concurrent model, we employ a Fourier basis of order 11, and for the historical influence model, we employ a Fourier basis of order 11 on both dimensions. The roughness parameter $\lambda$ is chosen using five-fold cross-validation. We consider seven time windows of length $\delta = 2, 4, 6, 10, 20, 30, \infty$ seconds for the historical influence model. Note when $\delta = \infty$, the model accounts for the causal effects from the beginning of the scan. When fitting the model for $Y$, combinations of window width for $Z$ and $M$ are implemented. Table 4 compares the mean squared prediction error (MSPE) of $M$ and $Y$ under different models when considering the STOP trial. From the table, the MPSEs of $M$ are very close across all types of models with the historical model using $\delta = 20$ as the lowest. The combination of $\delta_{YZ} = 6$ and $\delta_{YM} = 4$ yields the lowest MPSE of $Y$. Fixing $\delta_{YZ}$, $\delta_{YM} = 4$ achieves the lowest MPSE and $\delta_{YM} = \infty$ significantly higher than the rest. We choose the combination of $\delta_{MZ} = 20$, $\delta_{YZ} = 6$ and $\delta_{YM} = 4$ to estimate the causal direct and indirect trajectories. Considering the GO trials only, the estimated $\delta$ values are $\delta_{MZ} = 20$, $\delta_{YZ} = 4$, and $\delta_{YM} = 4$, which are very close to those under the STOP trials.

Our functional mediation models enable the estimate of causal curves for each individual. Figure 6 presents the estimated direct and indirect effect curves (as well as the 95% point-wise confidence band following the bootstrap procedure introduced in Section 3.2) from one of the 121 participants under the historical model with $\delta_{MZ} = 20$, $\delta_{YZ} = 6$ and $\delta_{YM} = 4$ for the STOP trial (Figure 6a) and $\delta_{MZ} = 20$, $\delta_{YZ} = 4$ and $\delta_{YM} = 4$ for the GO trial (Figure 6b). From the figure, for most of the STOP trials, there is a significant positive indirect effect; while under the GO trials, there exits more significant positive direct effect. Under the GO trial, we observe significant indirect effect at the beginning of the experimental session, which is not expected. One possible explanation is that the subject did not response correctly. Further investigation needs to be done to correlate the results with the behavior. With consecutive STOP trials, the indirect effects are accumulated and get stronger. These findings are consistent with the fact that preSMA functions as a crucial role in motion prohibition (Nachev et al., 2007), but from a dynamic perspective.

For comparison, an anterior preSMA region (MNI coordinate: (-4,36,56)) as the mediator is tested (Figure 7). The estimated influence window widths ($\delta$ values) are different under the STOP trial. From the figure, the indirect effect is not significant across the whole experimental session under the STOP trial and only significant at about 10–20 seconds un-der the GO trial. The same conclusion holds when applying the $\delta$ values estimated from above preSMA region (Figure B.1 in Supplementary Section B). This suggests that this brain region is not functioning on the movement involved in this response conflict task.

Table 4: Mean squared prediction error of $M$ and $Y$ under various types of functional mediation model under the STOP trial.

|   | Concurrent | Historical ($\sim M$) | Historical ($\sim Z$) | | | | | | | |
|---|---|---|---|---|---|---|---|---|---|---|
|   |   |   | $\delta = 2$ | $\delta = 4$ | $\delta = 6$ | $\delta = 10$ | $\delta = 20$ | $\delta = 30$ | $\delta = \infty$ |
| $M$ | 353.460 |   | 352.645 | 352.244 | 351.988 | 351.652 | **351.179** | 351.272 | 357.396 |
|   |   | $\delta = 2$ | 212.331 | 212.308 | 211.960 | 212.333 | 212.378 | 212.130 | 212.343 |
|   |   | $\delta = 4$ | 211.324 | 211.227 | **211.062** | 211.064 | 211.124 | 211.070 | 211.572 |
|   |   | $\delta = 6$ | 211.883 | 211.663 | 211.541 | 211.546 | 211.592 | 211.575 | 212.110 |
| $Y$ | 220.203 | $\delta = 10$ | 214.277 | 214.035 | 213.909 | 213.953 | 213.989 | 213.971 | 214.510 |
|   |   | $\delta = 20$ | 218.383 | 218.098 | 217.878 | 217.928 | 218.312 | 218.247 | 218.765 |
|   |   | $\delta = 30$ | 221.183 | 220.915 | 220.666 | 220.685 | 221.041 | 221.266 | 221.727 |
|   |   | $\delta = \infty$ | 295.291 | 294.938 | 294.904 | 294.695 | 294.820 | 294.742 | 301.385 |

## 6 Discussion

In this study, we introduce a functional mediation framework where the treatment, the mediator and the outcome are all functional measures. Two types of functional mediation models, (1) a concurrent mediation model and (2) a historical influence mediation model, are discussed. Causal estimands and identification assumptions are explored. Our framework allows the estimation of individual time-varying causal curves as the treatment trajectory may vary across subjects. As among the first attempt of functional mediation analysis, our framework is outlined under the setting of (sequentially) randomized treatment assignment regime and assuming no unmeasured confounding factors. Extensions to observational studies considering both baseline and time-varying covariates will be future directions of research.

No unmeasured "mediator-outcome confounder" is a strong assumption in practice (Imai et al., 2010), especially in fMRI studies when considering both mediator and outcome as brain activations (Fox et al., 2006; Mason et al., 2007; Obeso et al., 2013; Sobel and Lindquist, 2014). A next-step study will be considering the existence of unmeasured confounding. For scalar measures, the unmeasured confounding effect can be captured by the correlation between the two model error terms under the SEM framework (Imai et al., 2010; Zhao and Luo, 2014). The representation under functional data context may not be straightforward, and we leave the study of unmeasured confounding as well as sensitivity analysis to future research.

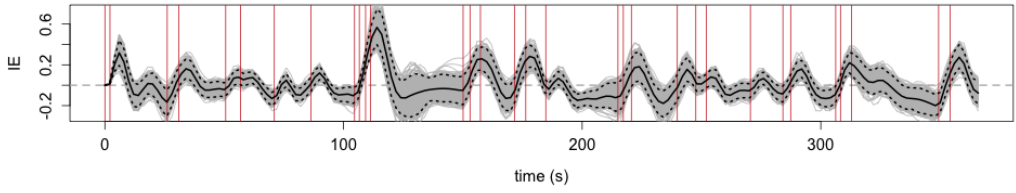
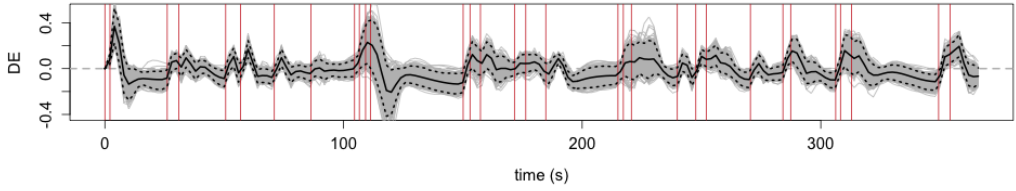

(a) STOP trial ($\delta_{MZ} = 20$, $\delta_{YZ} = 6$, $\delta_{YM} = 4$).

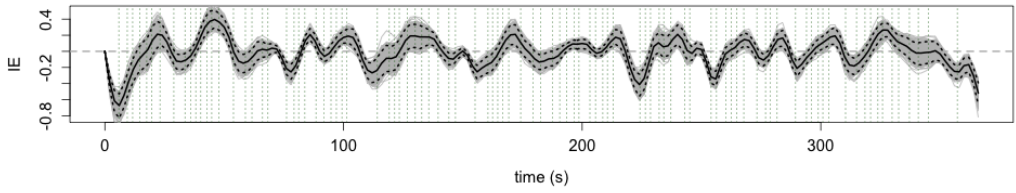
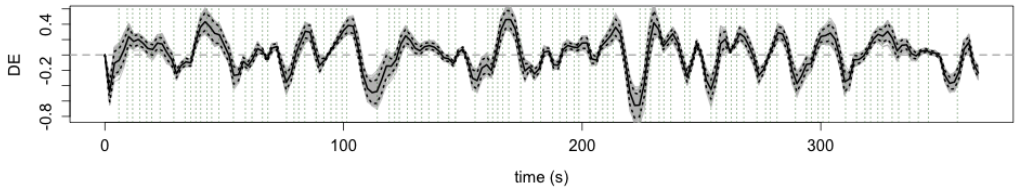

(b) GO trial ($\delta_{MZ} = 20$, $\delta_{YZ} = 4$, $\delta_{YM} = 4$)

Figure 6: Estimated indirect effect (IE) and direct effect (DE) curves (as well as point-wise 95% confidence interval from 500 bootstrap samples) of one participant under the historical mode for STOP and GO trials separately.

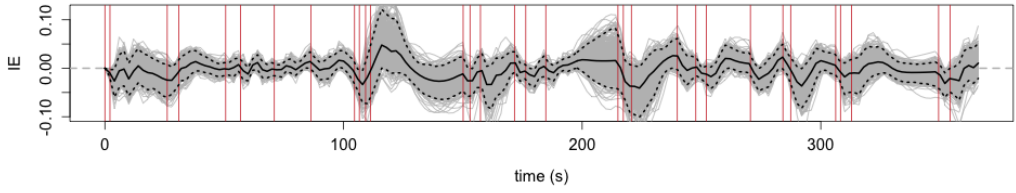

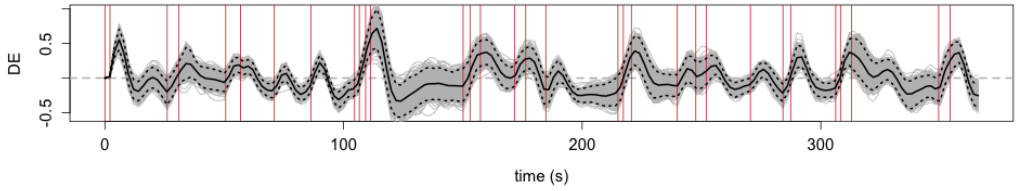

(a) STOP trial ($\delta_{MZ} = 8$, $\delta_{YZ} = 30$, $\delta_{YM} = 6$).

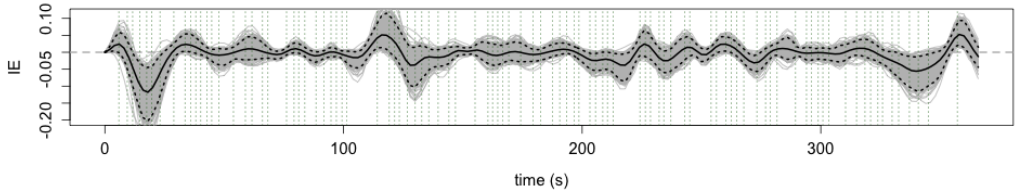

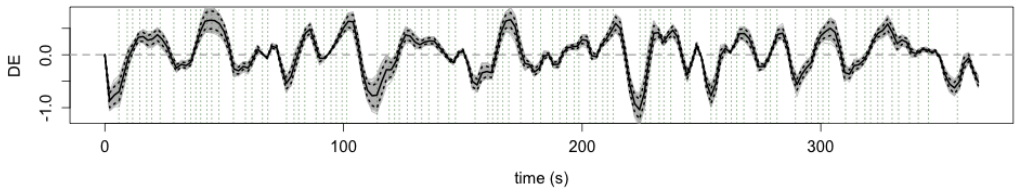

(b) GO trial ($\delta_{MZ} = 20$, $\delta_{YZ} = 4$, $\delta_{YM} = 6$)

Figure 7: Estimated indirect effect (IE) and direct effect (DE) curves (as well as point-wise 95% confidence interval from 500 bootstrap samples) of one participant under the historical model for STOP and GO trials separately with anterior preSMA (MNI coordinate: (-4,36,56)) as the mediator.

Applied to an fMRI data set, the proposed functional mediation analysis augments the current technologies for discovering dynamic brain connectivity. In our application, the same canonical HRF is employed on both brain regions. However, it has been shown that the HRF differs across brain regions as well as across individuals (Aguirre et al., 1998; Schacter et al., 1997; Vazquez et al., 2006). Considering different HRFs for the mediator and outcome brain regions, the treatment trajectories in the mediator and outcome models are consequently divergent though generated from the same stimuli onsets. Under this circumstance, a critical question is whether the assumption of stable treatment assignment regime is still valid. The exploration of "unstable" treatment trajectories will be a topic of future interest. Another drawback of the current application is in the design of the experiment. The STOP/GO trials are randomized, but the inter-trial interval is short creating difficulty in decomposing the HRFs between trials and the interpretation of the estimated influence window.

The BOLD signals in fMRI studies can be viewed as so-called "dense functional data" (Wang et al., 2016). Our framework can be applied to longitudinal studies as well, where the measures are considered as the "sparse functional data" and the parameteric $\sqrt{N}$ convergence rate cannot be attained. We are interested in extending the current framework to this type of data and studying the theoretical properties in the future.

## SUPPLEMENTARY MATERIAL

# A  Additional Simulation Results

Figure A.1 presents the mean squared prediction error (MSPE) of $M$ and $Y$ under different $\delta$ values in the simulation study when the true model is a historical mediation model with $\delta = 6$. From the figure, when the model is correctly specified, the MSPE yield the lowest. This simulation result suggests a way of choosing the width of influence window $\delta$.

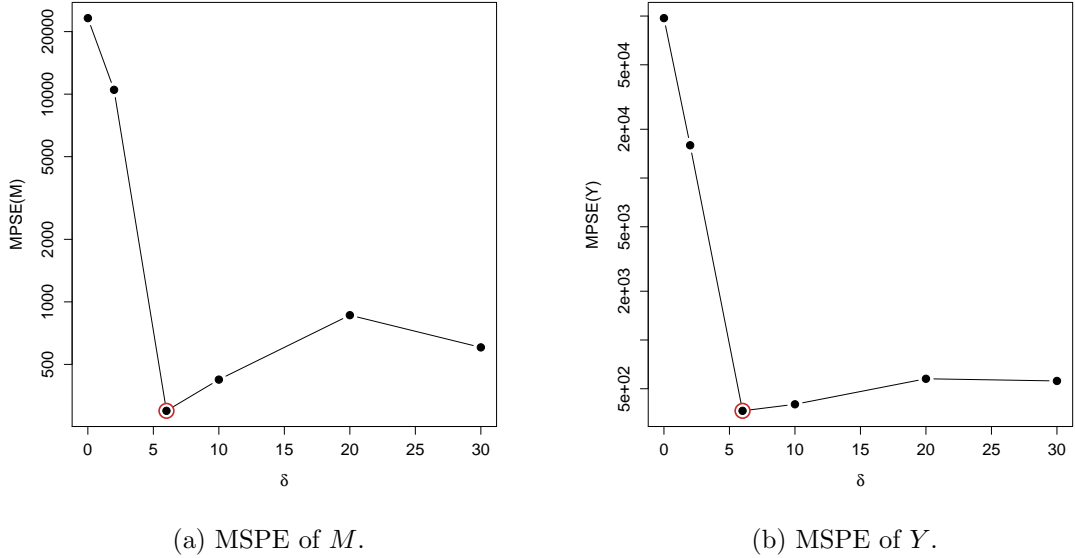

(a) MSPE of $M$.  (b) MSPE of $Y$.

Figure A.1: The mean squared prediction error (MSPE) of (a) $M$ and (b) $Y$ under different $\delta$ values when the true model is a historical mediation model with $\delta = 6$. $\delta = 0$ indicates the concurrent mediation model.

# B   Additional Functional MRI Study Results

Figure B.1 presents the estimated causal curves under the $\delta$ choices in Figure 6. From the figure, the estimated causal curves are slightly different from those in Figure 7, but the conclusion remains the same, i.e., there is no significant indirect or direct effect across time.

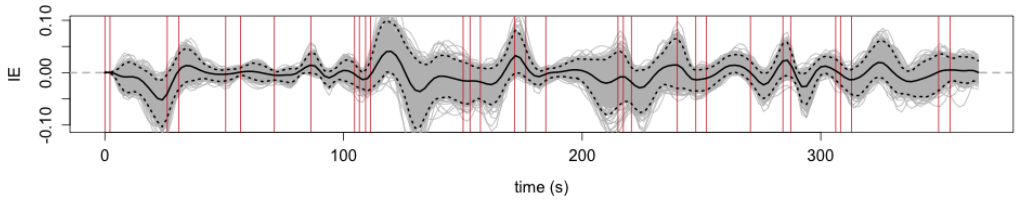
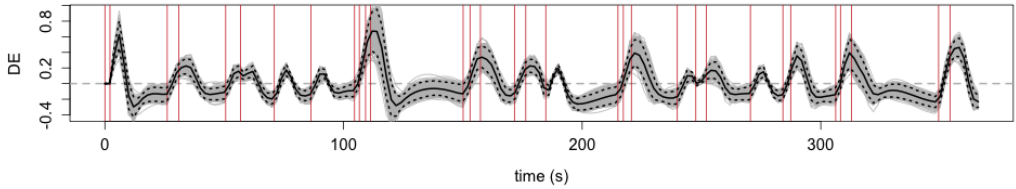

(a) STOP trial ($\delta_{MZ} = 20$, $\delta_{YZ} = 6$, $\delta_{YM} = 4$).

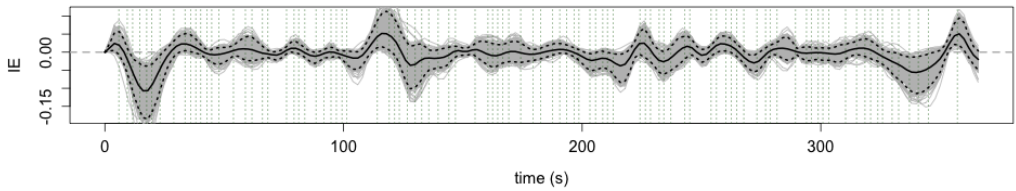
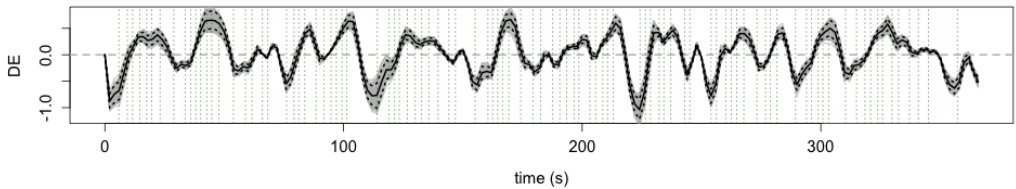

(b) GO trial ($\delta_{MZ} = 20$, $\delta_{YZ} = 4$, $\delta_{YM} = 4$)

Figure B.1: Estimated indirect effect (IE) and direct effect (DE) curves (as well as pointwise 95% confidence interval from 500 bootstrap samples) of one participant under the historical model for STOP and GO trials separately with anterior preSMA (MNI coordinate: (-4,36,56)) as the mediator. $\delta$ values are the same as in Figure 6.